\documentclass{ws-procs9x6}

\begin{document}

\title{Heavy Quark Free Energies and Screening in SU(2) Gauge Theory
\footnote{\uppercase{T}his work is supported by the 
\uppercase{DFG} \uppercase{G}rant No. \uppercase{FOR} 339/1-2.}}

\author{S. DIGAL, S. FORTUNATO, P. Petreczky}

\address{Fakult\"at f\"ur Physik\\ Universit\"at Bielefeld\\
33615 Bielefeld, Germany}


\maketitle

\abstracts{We investigate the singlet, triplet and
colour average heavy quark free energies in 
$SU(2)$ pure gauge theory at various temperatures $T$.
We focus on the long distance behaviour of the free energies,
studying in particular the temperature dependence of the string tension
and the screening masses.
The results are qualitatively similar to the $SU(3)$ scenario,
except near the critical temperature $T_c$ of the deconfining transition.
Finally we test a recently proposed method to renormalize the 
Polyakov loop.}

\section{Introduction}

         The interaction between quarks in a medium at some temperature
         $T$ has been subject of intensive investigations in the 
         last years~\cite{olaf}-\cite{felix}. 
         In general, one expects that the medium 
         strongly affects the interaction, especially at
         large separations $R$. Below the deconfinement temperature $T_c$
         the confining part of the interaction gets modified,
         so that the string tension $\sigma$ is a function of $T$.
         Above $T_c$ 
         the most important effect is the screening of the 
         colour charges, which leads to an exponentially 
         decreasing potential for $R\,{\gg}\,1/T$. 
         
         We investigate here the modification of interquark forces
         in terms of the free energy of a static quark-antiquark pair
         in $SU(2)$ pure gauge theory. We analyze the singlet, triplet
         and average colour channels. 
	 
	 Below $T_c$ we compare
         our data with the behaviour at $T=0$ and study the
         variation of the string tension with the temperature; above $T_c$ 
         we study the temperature dependence of the
         screening mass $\mu$ both for the average and for the
	 singlet free energies.

\section{Free energies in SU(2) gauge theory}

         On the lattice the free energy of the static quark-antiquark
         pair in the gluonic medium is determined by measuring 
         Wilson line correlation functions

          \begin{equation}
            e^{-F(R,T)/T+C}={\langle\,Tr\,L(\vec{R})Tr\,L^{\dagger}(\vec{0})\rangle}
          \end{equation}
          \begin{equation}
            e^{-F_{1}(R,T)/T+C}=2{\langle\,Tr\,(L(\vec{R})L^{\dagger}(\vec{0}))\rangle},
          \end{equation}
          \begin{equation}
            e^{-F_{3}(R,T)/T+C}=\frac{4}{3}\Big(e^{-F(R,T)/T+C}-\frac{1}{4}e^{-F_1(R,T)/T+C}\Big)
          \end{equation}

          where $F_1$, $F_3$ and $F$ refer to the colour singlet
          state, the colour triplet one
          and to the colour average of the free energy in the 
          singlet and adjoint channels, respectively. 
          The normalization constant $C$
          can be fixed in various ways, e.g. by 
          comparison with the $T=0$ free energies.
          In order to determine 
          the singlet and triplet free energies we fixed the
          Coulomb gauge, where both free energies coincide with
          a gauge independent definition which was recently proposed~\cite{owe}.        

	  The variation with $T$ of the
          free energy (especially the colour singlet one)  
          is also important because the corresponding
          potential could be used to calculate the
          binding energies of heavy mesons when $T{\neq}0$.
          These binding energies can be used to determine the 
          sequential suppression pattern of heavy quarkonia 
          as a function of the temperature~\cite{digal}.

\vspace*{-0.4cm}
\begin{figure}[htb]
\centerline{\epsfxsize=8.1cm\epsfbox{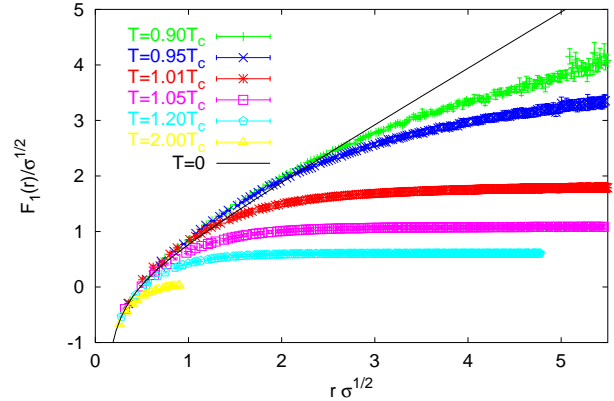}}   
\vspace*{-0.1cm}
\caption{Colour singlet free energy at various temperatures. \label{fig1}}
\end{figure}

          In Fig.~\ref{fig1} we show the behaviour of the colour singlet free energy.
	  The pattern looks the same as in $SU(3)$ gauge theory~\cite{felix}. 
	  The solid black line is the 
          fit to the $T=0$ potential done by the UKQCD collaboration~\cite{booth}.
          In the region $1<r\sqrt{\sigma}<2.5$ 
          we notice an enhancement compared
          to the $T=0$ curve (also found in~\cite{owe}). 

          In the deconfinement phase
          the curves reach a plateau at some distance
          $r_D$ which decreases with the temperature, as expected;
          near $T_c$ screening 
          sets in at distances $r\sqrt{\sigma}\,{\approx}\,1$, that is 
          of the order of 0.5 fm (assuming $\sqrt{\sigma}\,{\approx}\,400\,$MeV).

\section{Results Below $T_c$: String Tension}

          In the confinement phase the study of heavy-quark free energies allows
          to verify string models' predictions concerning the
          form of the potential~\cite{gao}. According to such predictions, at
          distances $R\gg{\frac{1}{T}}$, the potential 
          becomes 

          \begin{equation}
	     V(R,T)=V_0+\sigma(T)R+T\ln(2RT),
	  \label{eq4}
          \end{equation}

	  where the
	  logarithmic term is due to transverse fluctuations
	  of the string~\cite{luscher}.

\vspace*{-0.6cm}
	\begin{figure}[htb]	
        \centerline{\epsfxsize=7cm\epsfbox{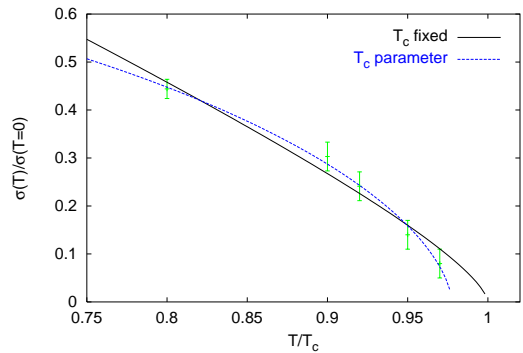}}   
	\vspace*{-0.2cm}
        \caption{String tension as a function of $T$ close to 
         the critical point. \label{sigT}}
	\end{figure}
	\vspace*{-0.2cm}

          We find that the
          singlet, triplet and average
          free energies relative to the same
          temperature converge to the same 
          line at large distances, 
          so that $\sigma(T)$ is identical
          no matter which colour channel one considers.
	  
	  By fitting the ansatz (\ref{eq4})
	  on the tail of the colour average curves, we derived the string tension $\sigma(T)$, which 
          we plot in Fig.~\ref{sigT} as a 
          function of $T/T_c$. 
          For temperatures very close to $T_c$, the fit
	  ansatz is no longer good because of finite size effects: this issue is still
	  under investigation.
          Near $T_c$ one expects that
          $\sigma(T)\,{\propto}\,(T_c-T)^{\nu}$, where
	  $\nu=0.63$ is the 3D Ising exponent. 
          We adopted the scaling ansatz used in~\cite{engels}, 
          $\sigma(T)=a(T_c-T)^{\nu}[1+b(T_c-T)^{1/2}]$, with $a$ and $b$ as free parameters. 
	  The solid
          line in the plot is the best fit curve, and we see that it is compatible
	  with the data. If we also take $T_c$ as a parameter of the fit, we get a much better
	  agreement for a curve centered at the
	 "pseudocritical" temperature $0.98T_c$ (dashed line in the plot). 
          
\vspace*{-0.1cm}

\section{Results Above $T_c$: Screening}

          At deconfinement, the determination of the free energies
          is important both to find the domain of validity
          of perturbation theory and to study the screening effects
          on the heavy quark free energy due to the medium of unbound coloured gluons.
          
\begin{figure}[htb]	
        \centerline{\epsfxsize=7.5cm\epsfbox{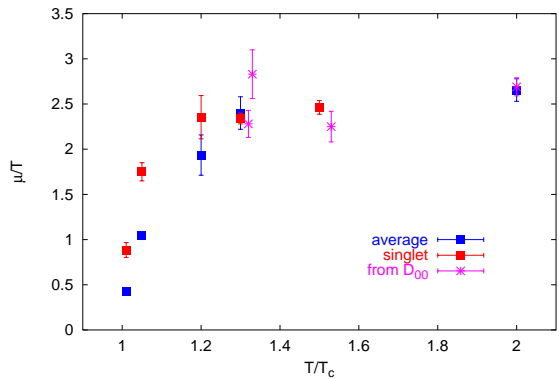}}   
	\vspace*{-0.2cm}
        \caption{Screening mass as a function 
          of the temperature. \label{scrmass}}
\end{figure}

          To extract the screening masses we applied the correlated fit introduced
          in~\cite{mckerr} to the singlet and average correlators. 
	  Our fit ansatz was 
          
          \begin{equation}
            \frac{F}{T}\,=\,\frac{A}{R^d}\exp(-\mu{R})+C
          \end{equation}
         
          where we considered $d=1,2$. Here we 
          set $C=-\ln\langle|L^2|\rangle$.
          The resulting values of the 
          screening mass are shown in Fig.~\ref{scrmass} as a function of $T/T_c$.

\vspace*{-0.4cm}
\vspace*{-0.2cm}

          In the plot the masses relative to the singlet 
          are the ones obtained from the fits with $d=1$: 
          our values are compatible with those obtained from the electric 
          propagator $D_{00}$ in Landau gauge~\cite{heller}. 
          Moreover we see that the mass decreases while approaching $T_c$, in contrast
          to the $SU(3)$ case~\cite{okacz}.
          For the average free energy and $T\,\approx\,T_c$ we obtain good fits only for
          $d=1$, whereas for $T\,\approx\,2T_c$ both 
          $d=1$ and $d=2$ yield reasonable $\chi^2$s. This gives a
          systematic error of about $20\%$ in the screening mass.

\vspace*{-0.1cm}
\section{Conclusions}

          We found that the $SU(2)$ free energies
          exhibit some peculiar features 
	  as compared with their $SU(3)$ counterparts, especially
	  near $T_c$,
          where the order of the deconfining transition plays an important role.
          Apart from the temperature
          dependence of the string tension, which
          was already known, we mention in particular the decrease of the electric screening 
          mass of the singlet free energy by approaching $T_c$.

\vspace*{-0.3cm}
	\begin{figure}[htb]	
        \centerline{\epsfxsize=7cm\epsfbox{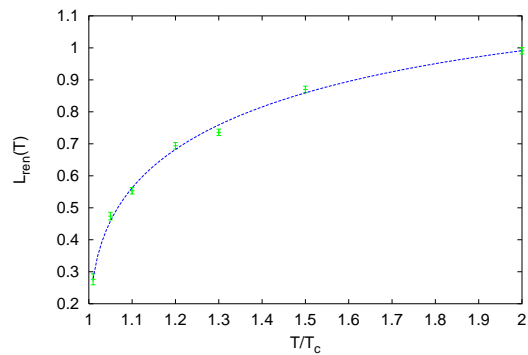}}   
\vspace*{-0.2cm}
        \caption{Renormalized Polyakov loop as a function of the
          temperature. We fitted the points with the ansatz
          $L_{ren}(T)=a(T-T_c)^{\beta}[1+b(T-T_c)^{1/2}]$, fixing $\beta=0.3265$.
          The best fit curve (dashed line) reproduces well the data near $T_c$.\label{repoly}}
	\end{figure}	

\vspace*{-0.2cm}

          Finally we show the pattern of the Polyakov loop (Fig.~\ref{repoly}), 
          renormalized by matching the free energy at short distances with the
          zero temperature heavy quark potential, as suggested in~\cite{felix}. 
          The points follow the expected behaviour,
          $L_{ren}~{\propto}~(T-T_c)^{\beta}$, where $\beta=0.3265$
          is the 3D Ising exponent.

\vspace*{-0.2cm}

\section*{Acknowledgments}

We would like to thank A. Cucchieri for giving us 
the gauge fixing program,  
R. D. Pisarski, J. Engels and O. Kaczmarek for stimulating discussions.

\end{document}